\documentclass[aps,superscriptaddress,amsmath,amssymb,showpacs,twocolumn]{revtex4}

\usepackage{hyperref}
\usepackage{epsfig}
\usepackage{color}

\newcommand{\ket}[1] {\left|#1\right\rangle}

\newcommand{\rem}[1]{}

\newcommand{\refe}[1]{~(\ref{#1})}

\begin{document}
\title{Detection of ultrafast oscillations in Superconducting Point-Contacts by means of Supercurrent Measurements}

\author{R. Avriller}
\affiliation{Donostia International Physics Center (DIPC), Paseo Manuel de Lardizabal 4,
E-20018 Donostia-San Sebastian, Spain.}
\affiliation{Laboratoire Ondes et Mati\`ere d'Aquitaine, UMR 5798, Universit\'e Bordeaux I and CNRS,
351 Cours de la Lib\'eration, F-33405 Talence Cedex, France}

\author{F. S. Bergeret}
\affiliation{Centro de F\'isica de Materiales (CFM-MPC), Centro Mixto CSIC-UPV/EHU,
Manuel de Lardizabal 5, E-20018 San Sebastian, Spain}
\affiliation{Donostia International Physics Center (DIPC), Paseo Manuel de Lardizabal 4,
E-20018 Donostia-San Sebastian, Spain.}

\author{F. Pistolesi}
\affiliation{Laboratoire Ondes et Mati\`ere d'Aquitaine, UMR 5798, Universit\'e Bordeaux I and CNRS,
351 Cours de la Lib\'eration, F-33405 Talence Cedex, France}

\date{\today}

\begin{abstract}
We present a microscopic calculation of the nondissipative current through a superconducting quantum
point contact coupled to a mechanical oscillator. 
Using the nonequilibrium Keldysh Green function approach, we determine the current-phase relation.
The latter shows that at certain phases, the current is
sharply suppressed. These dips in the current-phase relation provide information about
the oscillating frequency and coupling strength of the mechanical oscillator. 
We also present an effective two-level model from which we obtain analytical expressions
describing the position and width of the dips. Our findings are of relevance 
for nanomechanical resonators based on superconducting materials. 
\end{abstract}


\maketitle

\section{Introduction}
\label{sec1}

The observation of a small, albeit macroscopic, mechanical oscillator in the ground 
state \cite{oconnell:2010} constitutes a milestone on the road leading to the 
experimental verification of quantum decoherence in mechanical systems. 
Detection and manipulation of such nanomechanical oscillators are still a challenging
issue and several approaches are being pursued by experimentalists.
A nonexhaustive list of devices includes electromagnetic cavities in the microwave range \cite{regal:2008,Rocheleau:2009}, superconducting qubits \cite{suh:2010}, optical cavities 
\cite{arcizet:2006,Favero:2010,Aspelmeyer:2010}, single-electron transistors 
\cite{lassagne:2009,steele:2009}, and tunnel junctions \cite{flowers:2007}.
But in order to tackle fundamental questions of decoherence in macroscopic systems, \cite{blencowe:2004} still further improvements are necessary, thus motivating the investigation of new directions for the detection of nanoelectromechanical systems.

In Ref.\cite{flowers:2007} the modulation of the tunneling quantum amplitude 
in an atomic point contact has been exploited to detect the mechanical fluctuations 
of a doubly clamped beam. 
The current through the point contact is modulated by the change in the distance 
between the oscillating beam and a fixed reference metal. 
It has been shown that this kind of detector can reach the quantum 
limit of displacement detection \cite{bocko:1996,clerk:2004} and can be allowed, 
in the experiment of Ref.\cite{flowers:2007}, to measure with a good 
accuracy the resonating frequency of the oscillator and its Brownian motion. 

It seems feasible to reproduce a similar experiment with superconducting 
leads instead of normal metal ones. 
In this case the atomic point contact forms a Josephson 
junction between the mobile and fixed leads. 
Josephson current in similar tunnel junctions has been demonstrated 
by using a superconducting scanning tunneling microscope (STM) 
tip \cite{naaman:2001}. Also the current phase relation has been measured in Josephson junctions
consisting of atomic point contacts \cite{DellaRocca2007,Zgirski2011} and carbon nanotubes \cite{Pillet2010}. 
An alternative system with a similar modulation of the tunneling amplitude
induced by a mechanical displacement is a suspended carbon nanotube contacted 
between two superconductors in the Fabry-Perot regime.

From the theoretical point of view,
the effect of the modulation of the tunneling matrix element due to 
mechanical oscillations in Josephson junctions has been considered 
in the literature, but only in the adiabatic limit of slow 
variation of the tunneling amplitude \cite{zhu:2006,fransson:2008}. 
This limit is adapted to the tunneling case where the only 
relevant energy scale is the superconducting gap $\Delta$. 
Thus for oscillating frequencies much smaller than 
$\Delta/\hbar$ (typically of the order of tens of GHz) 
the time dependence of the oscillator can
be treated adiabatically with a correction proportional to the 
time derivative of the displacement \cite{zhu:2006}. 

The situation is different for an atomic point contact consisting of
few conducting channels, some of them with high transmission.  
It is well know \cite{furusaki,Beenakker1991} 
that in such junctions the supercurrent 
is controlled by the occupation of the Andreev bound states (ABSs) with energies depending on the transmission
and phase difference across the contact. In the case of a unique conducting channel,  
there are only two Andreev states with energy
\begin{equation}
\pm \omega_{A}(\phi)=\pm\Delta \sqrt{1-\tau \sin^{2}(\phi/2)}\; ,
\label{abs}
\end{equation}
where $\phi$ is the phase difference between the two superconductors,
and $\tau$ ($0\leq \tau \leq 1$) the transmission coefficient of the channel.
For $\tau\ll 1$ (tunneling limit) the spectrum of the two-level system as a function 
of the phase difference is almost flat with a level spacing between the channels equal to $2\Delta$. 
In the opposite limit, for $\tau = 1$, the level spacing between the ABSs depends on $\phi$ and
shows a minimum at $\phi= \pi$ where the energy splitting vanishes. 
Thus,  
the energy level splitting $2\omega_A$ spans the range $2\Delta\geq2\omega_A\geq 2\Delta \sqrt{1-\tau}$ and may be equal to the mechanical resonating frequency.  
Moreover, for large enough phases  the two Andreev levels can be  deep inside the superconducting gap and an
effective two-level model description for the contact that neglects the continuum part of the spectrum can be used \cite{Zazunov2003}.  

A two-level system is the simplest example of a quantum detector \cite{ClerkRMP}.
By tuning the energy splitting between the two levels to a radial frequency $\omega_{0}$
one can measure the transition rates between the two levels induced 
by the coupling to an external system and  
thereby determine the fluctuation spectrum at $\omega_0$. 
For the two-level system formed by the Andreev states it has been
predicted very recently \cite{bergeret:2010,bergeret:2011} that under microwave irradiation 
of radial frequency 
$\omega_{0}$ the current-phase relation of the Josephson junction show 
dips at values of $\phi$ that are solutions of the 
equation $\hbar \omega_{0}=2 \omega_{A}(\phi)$.
At these values of the phase the electromagnetic field induces
transitions between the two Andreev levels and the resulting current 
vanishes at resonance.
The question thus naturally arises if a similar mechanism can be 
used to detect mechanical oscillations. 

In this paper we explore the effect of 
fast modulation of the transparency of a single channel quantum
point contact on the current-phase relation (CPR). The origin of that modulation is  
the vibration of the contact, with oscillations in the range of hundreds of MHz.
We show that if the frequency of the mechanical oscillator is of the order of the 
spacing between the Andreev levels, then the CPR differs drastically from the one obtained in the adiabatic case. 
In analogy with the microwave irradiation, we find that the 
modulation of the tunnel amplitude leads also to the appearance 
of dips in the current-phase relation that, for 
typical experimental situations, could be extremely 
sharp.
Measurement of the current phase relation could thus be 
used to detect the resonating frequency of the 
mechanical oscillator.
In order to describe this effect it is clear that one has to go beyond the adiabatic 
approach of Ref. \cite{zhu:2006} in order to take into account the transitions between ABSs.

The plan of the paper is the following.
In Sec. \ref{sec2}, we first 
introduce a microscopic Hamiltonian to describe the superconducting 
point contact in the presence of electromechanical vibrations. From this 
model Hamiltonian, we compute the supercurrent through the junction in 
two ways. In a first approach, Sec. \ref{subsec3-1}, we use an effective 
two-level model similar to the one derived in Ref. \cite{Zazunov2003}.
Within this model, we derive analytical expressions for the current and for the 
occupation of the Andreev states. Second, in Sec. \ref{subsec3-2}, we  
calculate the supercurrent with the help of the Keldysh Green
functions and check the validity of the two-level model. In particular 
we  present  numerical solutions for the current phase relation.
In Sec. \ref{sec4-2} we discuss the use of this method to detect mechanical 
oscillations and present the conclusions in Sec. \ref{sec5}.

\section{Model}

\label{sec2}

\begin{figure}[ht]
\centering
  \includegraphics[width=0.35\textwidth]{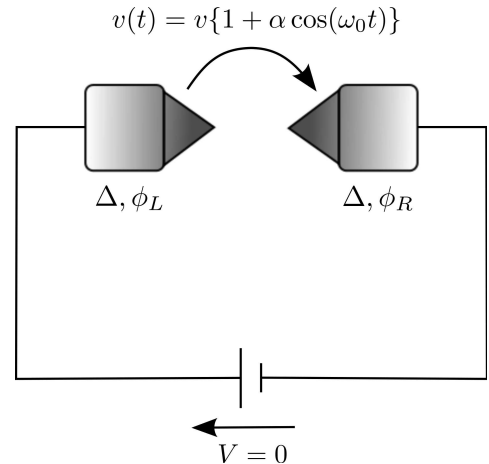}
\caption{\label{Model:fig} 
Schematic representation of a Josephson junction.
We assume that the distance between the two leads is modulated periodically at 
a frequency $\omega_{0}$. This induces a weak modulation of the quantum
hopping matrix element $v(t)$. 
}
\end{figure}

We consider a model for a Josephson junction where the normal state electronic hopping term
is modulated in time at a given frequency.
An example of the system considered is depicted in Fig.~\ref{Model:fig}. 
%
%
It consists of two bulk superconductors, characterized by a BCS gap $\Delta$, connected by a 
junction whose dimensions are assumed much smaller than the superconducting coherence length. 
This model can describe, for example, the contact between a STM superconducting tip and a bulk superconductor, 
or a quantum point contact in break junctions \cite{DellaRocca2007}. 
The same model can also describe a quite different system: a suspended carbon nanotube between 
superconductors in the Fabry-Perot regime for a transparent barrier \cite{kong:2001}.
Different groups have realized suspended carbon nanotubes with good mechanical properties \cite{lassagne:2009,steele:2009}.
Also transparent contacts between superconductors and nanotubes forming SQUIDs have been 
observed \cite{Cleuziou:2006,Pillet2010}.
The motion of the nanotube induces a modulation of the gate potential seen 
by the nanotube and 
thus modulates the transparency of the electronic mode. 
Note also that the mechanical coupling considered here is different from the one investigated in 
Ref. \cite{Sonne:2010} for a suspended carbon nanotube in presence of magnetic field.

We will assume that the phase difference between the left (L) and right (R) superconducting leads, 
$\phi=\phi_{L}-\phi_{R}$ is time independent, i.e. there is no voltage drop at the junction. 
We consider the case of a single-channel superconducting point contact which can be described, for instance, 
by the following tight-binding Hamiltonian \cite{Alfredo:1995} (we set the units $\hbar=e=k_{B}=1$)
\begin{eqnarray}
\hat{H}(t) = \sum_{X=L,R}\hat{H}_{X} + \hat{V}_{EM}(t)\; .
\label{eqn:2}  
\end{eqnarray}
Here $\hat{H}_{X=L,R}$ are the Hamiltonians of the isolated $X=L,R$ leads given by 
\begin{equation}
\hat{H}_{X}(t) = \sum_{i\in X} \psi_{i}^{\dagger} \Delta \hat{\sigma}_{x} \psi_{i} +
\sum_{<i,j> \in X} \big{\lbrace} \psi_{i}^{\dagger} \hat{v}_{0}\psi_{j} + \mbox{H.c} \big{\rbrace}\; ,\label{eqn:3}
\end{equation}
where $\hat{v}_{0}$ is the hopping matrix between next-nearest-neighbor sites within
the isolated $X=L,R$ superconductors. We use the standard notation for the
Pauli matrices in Nambu space $\lbrace \hat{\sigma}_{x},\hat{\sigma}_{y},\hat{\sigma}_{z} \rbrace$.
The second term in the r.h.s. of Eq.\refe{eqn:2} describes the (time dependent) hopping between the leads defined as
\begin{eqnarray}
\hat{V}_{EM}(t) &=& \psi_{L}^{\dagger} \hat{v}(t)\psi_{R} + \psi_{R}^{\dagger} \hat{v}^{\dagger}(t)\psi_{L}   
\label{eqn:4} \\
\hat{v}(t) &=& v(t)\hat{\sigma}_{z} e^{i\phi \hat{\sigma}_{z}/2}
\,.  
\label{eqn:5} 
\end{eqnarray}
Notice that by a standard choice of the gauge, 
we have included the superconducting phase difference into
the hopping term $\hat{v}$.

We focus our study on the electromechanical (EM) properties of the junction. 
We assume that one of the leads is vibrating at a radial frequency 
$\omega_{0}$, thus modulating the hopping term between the left and the right lead. 
In the limit where the amplitude of
the oscillations is small, the time-dependent hopping will be linearly modulated as
\begin{eqnarray}
v(t) = v\big{\lbrace} 1 + \alpha \cos(\omega_{0}t) \big{\rbrace}\; ,
\label{eqn:1}  
\end{eqnarray}
where $\alpha\equiv a {(1/ v)}{dv/d x} \ll 1 $, with $x$ the displacement 
of the oscillating lead and $a$ the amplitude of oscillation. 
We will give estimations for $\alpha$ in Sec. \ref{sec4-2}.

For large amplitude oscillation a detailed microscopic model 
of electron transport is needed. The standard tunneling picture gives 
a simple exponential dependence of the hopping on the displacement
$v(x)=v(x_0) e^{-(x-x_0)/\lambda}$ where $\lambda$ is the tunneling length.
But this picture would be different for the suspended nanotube, where a 
linear dependence on the displacement is supposed to hold for large 
oscillation amplitudes. 
For this reason in this paper we concentrate on the linear displacement model,
which demands only a single parameter ( $\alpha$ ) to describe the EM coupling.

In the present model, the vibrational state of the junction is considered as an external time-dependent
perturbation, without its own dynamics \footnote{The question of the feedback of the vibrational dynamics
of the junction on the Josephson current is beyond the scope of the present paper}.  
Before proceeding to determine the current from the microscopic Hamiltonian [Eq.\refe{eqn:2}],
we start the next section by deriving an expression for the Josephson current within the framework of an
effective two-level Hamiltonian. 
We will see that in a certain range of parameters, such a model provides an accurate description
of the electronic dynamics of the contact.

\section{dc Josephson current}

\subsection{Andreev Two-Level Model}
\label{subsec3-1}

In a single-channel superconducting junction, as the one described by 
Eq.\refe{eqn:2},
the equilibrium spectral density is characterized by the two ABSs $\ket{+}$ and $\ket{-}$ with energies given in Eq.\refe{abs}. 
In terms of the hopping [Eq.\refe{eqn:5}], the transmission factor $\tau$ of the junction
is defined as
\begin{equation}
 \tau=\frac{4\beta}{(1+\beta)^2}\; ,
 \label{tcoeff}
\end{equation}
where $\beta=(v/v_{0})^2$ is the ratio between the tunnel hopping amplitude
and the electrode bandwidth $v_{0}$.
In the equilibrium case, the dc current is carried exclusively by the ABSs and can be written as the sum of
two opposite contributions \cite{Beenakker1991}
\begin{eqnarray}
I_{DC}^{(0)} &=& I_{-}n_{-} + I_{+}n_{+} \,,
\label{eqn:6} \\
I_{-} &=& - I_{+} = - 2 \frac{\partial}{\partial \phi} \omega_{A}
= \frac{\Delta^2 \tau \sin(\phi)}{2\omega_{A}}  
\label{eqn:7}\; ,
\end{eqnarray} 
where $n_{\pm}$ is the occupation of the $\ket{\pm}$ 
ABSs which is given by the Fermi distribution function $f(\pm\omega_{A})$.
Thus, in the finite temperature $T$ case one finally obtains the well known expression
for the Josephson current (see Appendix)
\begin{eqnarray}
I_{DC}^{(0)}(\phi) = \frac{\Delta^2 \tau \sin(\phi)}{2\omega_{A}}\tanh(\frac{\omega_{A}}{2 T}) 
\, .
\label{eqn:8} 
\end{eqnarray}  
In the limit of zero temperature, only the 
negative ABS ($\ket{-}$) is populated,
and contributes positively to the current in Eq.\refe{eqn:8}. 
At finite temperature $T$, the positive ABS ($\ket{+}$) gets populated due to the 
thermal smearing of the Fermi distribution and, according to Eq.\refe{eqn:6} contributes negatively
to the Josephson current.

Now let us consider the perturbation originated by the mechanical oscillations.
If the frequency and amplitude of the perturbation are  sufficiently small, one can still describe the current
as the contribution of the two ABSs. 
In this case it is convenient to work with an effective two-level model
Hamiltonian similar to the one derived in Refs.~\cite{Zazunov2003,Zazunov2005} 
from the microscopic Hamiltonian Eq.\refe{eqn:2}.
The main difference between the problem at hand and that considered in Refs.~\cite{Zazunov2003,Zazunov2005}
is that the time-dependent parameter is not the superconducting phase, but the transparency 
of the junction.
Adapting the method to our problem, it gives the following effective time-dependent Hamiltonian
for $d\tau/dt \ll \Delta/\hbar$:
\begin{equation}
 \hat h(t)= \Delta \cos\frac{\phi}{2}\hat\sigma_z+\Delta\sqrt{1-\tau(t)}\sin\frac{\phi}{2}\hat\sigma_y\; .
 \label{tlhb}
\end{equation}
This Hamiltonian is written in the ballistic basis of right and left moving electrons that is the eigenbasis
in the perfectly transmitting case ($\tau=1$). However, it is more convenient to write the two-level
Hamiltonian in the instantaneous Andreev basis \cite{bergeret:2011}.
For that sake, one performs a time-dependent unitary transformation
$\hat{H}=\hat U^\dagger \hat h \hat U-i \hat U^\dagger d\hat U/dt$,
where 
$\hat U(t)=
e^{-i\hat\sigma_z\frac{\pi}{4}}e^{-i\theta(t)\hat\sigma_y}$ and $\theta(t)=(1/2)\arctan[\sqrt{1-\tau(t)}\tan\phi/2]$.
The two-level Hamiltonian in the instantaneous Andreev basis is then given by
\begin{equation}
\hat{H}_A=\omega_A\hat \sigma_z+\frac{1}{8}
\frac{d\tau}{dt}
\frac{1}{\sqrt{1-\tau(t)}}\frac{\sin\phi}{1-\tau(t)\sin^2(\phi/2)}\hat\sigma_y \label{tlha}
\,.
\end{equation}
The off-diagonal terms describe the coupling between the Andreev levels
due to the EM oscillations. The current operator written in the same basis reads 
\begin{equation}
 \hat{I}_A=2 \frac{\partial \omega_A}{\partial \phi}\hat \sigma_z+\frac{\Delta^2\sqrt{1-\tau(t)}}{\omega_A}\hat\sigma_x \label{tlia}\; .
\end{equation}
We consider here the time-dependent transmission determined by expressions Eqs. (\ref{eqn:1}) and (\ref{tcoeff}) 
which in a linear approximation with respect to the amplitude $\alpha$ is
\begin{equation}
 \tau(t) \approx \tau + 2\tau\sqrt{1-\tau}\alpha\cos(\omega_{0}t) \, .
\label{linearEq}
\end{equation}
%
%
%
In order to avoid unphysical values of $\tau$ larger than one one has to impose $\alpha<\sqrt{1-\tau}/2\tau$. More over in the particularly interesting case of  very transparent 
channel,  the linear term of Eq.\refe{linearEq} vanishes and 
it would be  necessary to consider the quadratic one.  Thus, for ($\tau \rightarrow 1$)
comparison between the linear and quadratic term imposes the condition 
$\alpha \ll 2\sqrt{1-\tau}$.

We can now apply the method developed in Ref. \cite{bergeret:2011} to obtain time-averaged 
quantities like the current and the level population 
close to the first resonance, {\it i.e.} $\omega_{0}\approx2\omega_A$.
Within the rotating-wave approximation and for one-phonon assisted
processes we obtain for the dc current
\begin{eqnarray}
  I_{DC} \approx I_{DC}^{(0)}
\left[ 1 -
\frac{\Omega_{R}^{2}}{(2\omega_{A}-\omega_{0})^{2}+\Omega_{R}^{2}} \right] \, , 
\label{eqn:11} 
\end{eqnarray}   
where the Rabi frequency 
\begin{equation}
\Omega_{R} = \alpha \tau \sin(\phi)\omega_{0}\Delta^{2}/(2\omega_{A})^{2}
\label{rabi}
\end{equation}
is proportional to the coupling strength $\alpha$. 
Notice that at the resonance, the dc current vanishes.
This is due to the fact that when $\omega_0=2\omega_A$, resonant transitions between the Andreev levels take place
and both levels will be on average equally populated.
According to Eq.\refe{eqn:6}, this leads to a decrease of the dc current, which vanishes exactly at the resonance.
The width of the resonance as given by Eq.\refe{rabi} is proportional to the amplitude $\alpha$ of the oscillation.
Thus, in principle by measuring the CPR one could determine both the amplitude and frequency of the contact
oscillation as given by Eqs.~(\ref{eqn:11}) and  (\ref{rabi}).
Finally, the time averaged population of the upper and lower Andreev states are given by 
\begin{eqnarray}
n_{-} &=& 1 - \frac{1}{2}\frac{\Omega_{R}^{2}}{(2\omega_{A}-\omega_{0})^{2}+\Omega_{R}^{2}} \label{eqn:9} \,, \\
n_{+} &=& \frac{1}{2}\frac{\Omega_{R}^{2}}{(2\omega_{A}-\omega_{0})^{2}+\Omega_{R}^{2}} \,.
 \label{eqn:10} 
\end{eqnarray}  

These results are particularly simple and transparent, but they rely on different approximations.
In the following section we will thus perform a fully microscopic calculation in order to 
check their validity.
The method used does not need a hypothesis on the 
slowness of the mechanical frequency nor on the 
amplitude of the oscillations ($\alpha$) and 
keeps the full description of the electronic system.
The approach allows us to take into account the effect of the continuum spectrum
and in particular will remove the limitation on $\alpha \ll \sqrt{1-\tau}$.
We will see that the largest deviations are exactly where 
higher orders of $\alpha$ become important ($\alpha>\sqrt{1-\tau}$)
and when the resonant $\phi$ is close to $\pi$.

\subsection{Nambu-Keldysh Green function method}
\label{subsec3-2}

\begin{figure}[ht]
\centering
  \includegraphics[width=0.5\textwidth]{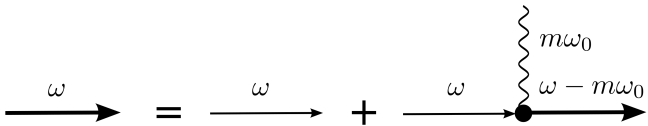}
\caption{\label{Dyson:fig} 
Schematic representation of the Dyson equation [Eq.\refe{dyson}] for the Green function in the 
Electrode-Nambu-Keldysh-Floquet space. The thick (thin) lines represent the Green functions in the vibrating
(nonvibrating) case, while the wavy line represents the external time-dependent perturbation.
}
\end{figure}

In the preceding section we have determined the dc Josephson current from an effective two-level model. 
In this section we introduce a numerical method that allows us to compute exactly
the dc Josephson current in the presence of EM coupling from the microscopic model introduced in 
Sec. \ref{sec2}. 
From our results, we will able to verify the range of validity of Eqs. (\ref{eqn:11}) and (\ref{rabi}),
and to obtain the current out of that range.
The method is based on the computation of
the Nambu-Keldysh Green functions (GFs) for the fermionic fields of the Hamiltonian \refe{eqn:2}.  
These are defined as
\begin{eqnarray}
\hat{G}_{XX'}^{\alpha\beta}(t,t') = - i \Big{\langle} T_{c} \psi^{\phantom\dag}_{X}(t) \psi_{X'}^{\dag}(t')\Big{\rangle}
\,,
\label{eqn:12} 
\end{eqnarray} 
where $T_{c}$ means time ordering along the Keldysh contour, $\alpha,\beta=\pm$ denote the Keldysh branches,
and $X,X'=L,R$ stand for electrode indexes.
From charge conservation at the L-R interface, one obtains the expression of the mean current crossing the junction
at time $t$ in terms of the GFs
\begin{eqnarray}
I(t) = \mbox{tr}\Big{\lbrace} \hat{\sigma}_{z} \big{\lbrack} \hat{v}(t)\hat{G}_{RL}^{+-}(t,t)
- \hat{v}^{\dagger}(t)\hat{G}_{LR}^{+-}(t,t) \big{\rbrack} \Big{\rbrace}
\,.
\label{eqn:13} 
\end{eqnarray} 
It is convenient to rewrite the Hamiltonian [Eq.\refe{eqn:2}] as the sum of an unperturbed part plus
the time-dependent perturbation
\begin{equation}
\hat H=\hat{H}_{0} + \hat{V}(t) \label{hopht}\;,
\end{equation}
where
\begin{eqnarray}
\hat H_0 &=& \sum_{X=L,R}\hat H_X + v \hat W_T \, ,\\
\hat V(t)    & =& \alpha v \cos(\omega_0 t) \hat W_T \,,
\end{eqnarray}
and
\begin{equation}
\hat W_T =  \psi_{L}^{\dagger} \hat{\sigma}_{z} e^{i\phi \hat{\sigma}_{z} /2}\psi_{R} 
	+ 
	\psi_{R}^{\dagger} \hat{\sigma}_{z} 
e^{-i\phi  \hat{\sigma}_{z} /2}\psi_{L}
\,.
\end{equation}
The corresponding Dyson equation is shown diagrammatically in Fig.~\ref{Dyson:fig}.
%
%
%
%
%
The thin lines represent the unperturbed Green's functions, i.e., those associated to $\hat H_0$, while the thick
lines are the exact GFs which take into account the external time-dependent perturbation represented by a wavy line. 
\noindent
Due to the time periodicity of the 
Hamiltonian one can write the GFs in Floquet representation
and the corresponding current operator as a Fourier series 
\begin{eqnarray}
&&\hat{G}_{XX'}^{\alpha\beta}(t,t') = \sum_{n}e^{-in\omega_{0}t'}\int\frac{d\omega}{2\pi}
e^{-i\omega(t-t')}\hat{G}_{XX';n}^{\alpha\beta}(\omega) \, , \nonumber \\
&& \label{eqn:14} 
\\
&&I(t) = \sum_{n}e^{-in\omega_{0}t}I_{n} \,.
\label{eqn:15} 
\end{eqnarray}  
We are interested in the dc component ($n=0$) of the mean current
given by 
\begin{eqnarray}
\lefteqn{I_{DC}(\phi) =}&&\nonumber\\
 && \sum_{m} \int\frac{d\omega}{2\pi}\mbox{tr}\Big{\lbrace} \hat{\sigma}_{z} \big{\lbrack} 
\hat{v}_{m0}\hat{G}_{RL;0m}^{+-}(\omega) 
- \hat{v}^{\dagger}_{m0}\hat{G}_{LR;0m}^{+-}(\omega) \big{\rbrack} \Big{\rbrace}
\,,
\nonumber \\ && 
\label{eqn:16}
\end{eqnarray}
where the hopping matrix $\hat v$ in the Floquet space is defined as
\begin{equation}
\hat{v}_{m0} =\hat{v} \big{\lbrace} \delta_{m,0} + \frac{\alpha}{2} \delta_{m,\mp 1} \big{\rbrace} \,.
\label{eqn:17}
\end{equation} 
The Dyson equation in the frequency representation is then given by (cf. Fig.~\ref{Dyson:fig})
\begin{eqnarray}
\lefteqn{\hat{G}_{XX';n}^{\alpha\beta}(\omega) = \hat{g}_{XX'}^{\alpha\beta}(\omega)\delta_{n,0} +}
\nonumber \\
&&
\sum_{X_1,\alpha_1,m}
\left[ 
\hat{g}_{XX_{1}}^{\alpha\alpha_{1}}(\omega)\hat{V}_{X_{1}\overline{X}_{1};m}^{\alpha_{1}}
\hat{G}_{\overline{X}_{1}X';n-m}^{\alpha_{1}\beta}(\omega-m\omega_{0}) 
\right] \; .
\nonumber\\
&& 
\label{dyson}
\end{eqnarray}
Here the self-energy term is given by
\begin{equation}
\hat{V}_{X_{1}\overline{X}_{1};m}^{\alpha_{1}} =\alpha_{1}\big{(}\frac{\alpha}{2}\big{)}\hat{v}_{X_{1}\overline{X}_{1}}\delta_{m;\pm 1} \,,
\label{eqn:19} 
\end{equation} 
and the expressions for the free propagators $\hat{g}_{XX'}^{\alpha\beta}(\omega)$ in the absence of vibrations ($\alpha=0$)
are given in the Appendix.

In the case $\alpha \ne 0$, however, the GFs $\hat{G}_{XX';n}^{\alpha\beta}(\omega)$ can only be found
numerically by solving the Dyson equation (\ref{dyson}). The latter constitutes a linear system in the
Electrode-Nambu-Keldysh-Floquet space of dimension
$2 \times 2 \times 2 \times (2N_{ph}+1)$ and is solved by exact numerical inversion.
The maximum number of vibrational quanta in the system $N_{ph}$ is increased until convergence of the solution
is found \footnote{Typically, for moderate values of the
coupling strength ($\alpha \le 0.1$), it sufficient for the result to converge to choose $N_{ph}=3$.}.


\section{Discussion of the results}
\label{sec4-2}

\begin{figure}[ht]
\centering
  \includegraphics[width=0.5\textwidth]{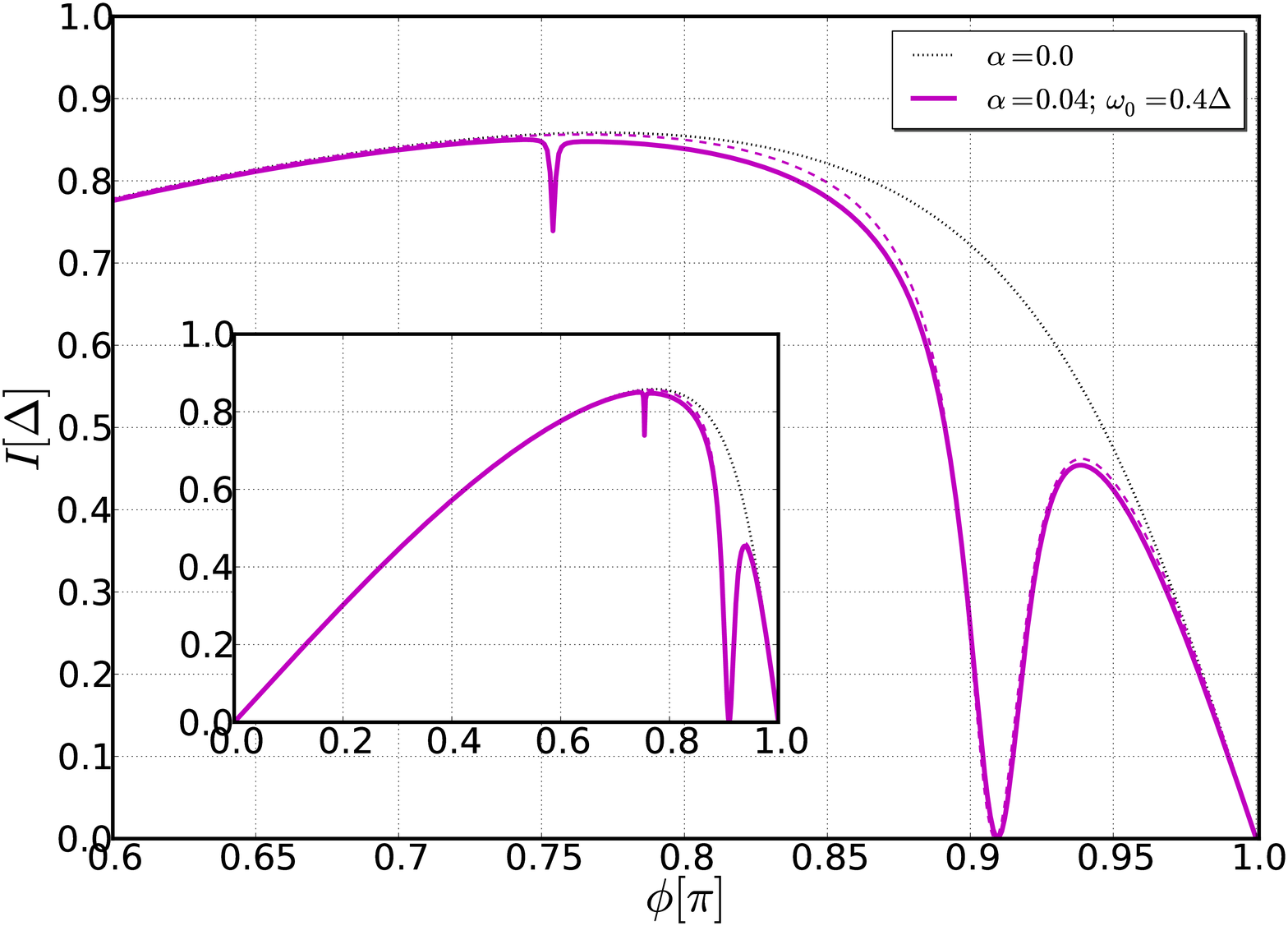}
\caption{\label{JJ_Current:fig} 
Josephson current for $\omega_{0} = 0.4\Delta$, $\tau=0.98$, $\alpha=0.04$, $T=0\mbox{ K}$, and $\eta=0.001\Delta$ ($\eta$ is the inverse  of the 
relaxation time of the superconductor as defined in the Appendix). 
Plain (dashed) curves correspond to the 
full numerical (analytical) calculations.
Dotted curve: current in the case of vanishing $\alpha$.
Inset: Same plot on a larger scale.
}
\end{figure}

\begin{figure}[ht]
\centering
  \includegraphics[width=0.5\textwidth]{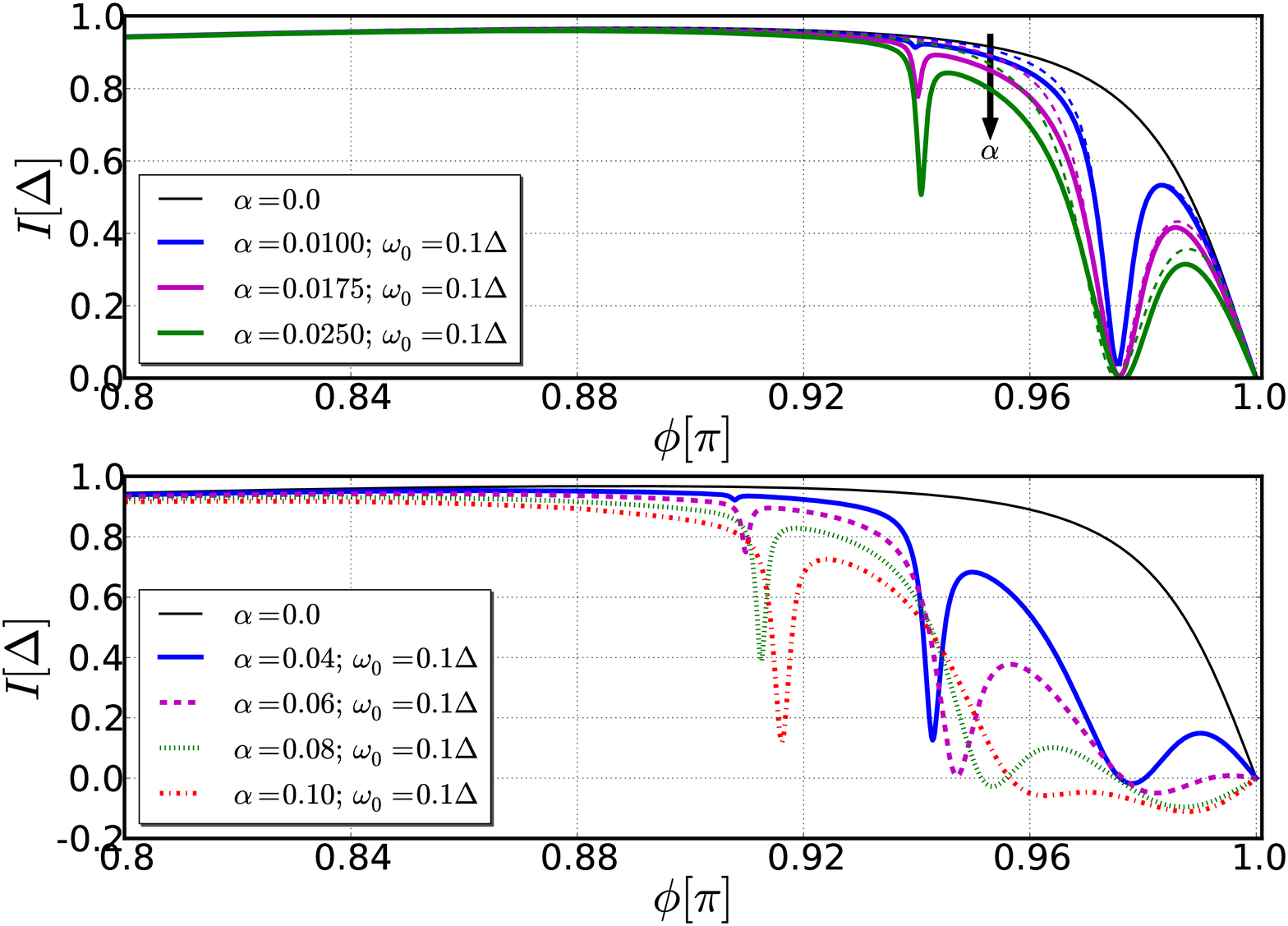}
\caption{\label{JJ_Current_Realistic_Parameters:fig} 
Upper panel: Josephson current for the cases of EM coupling $\alpha = 0.01,0.0175,0.025$ 
(within the range of validity of the analytic model).
Plain (dashed) curves correspond to the full numerical (analytical) calculation.
The arrow is denoting the direction of increasing values of $\alpha$. 
Lower panel: Josephson current for the cases of larger 
EM coupling $\alpha = 0.04,0.06,0.08,0.1$, as obtained from the full numerical calculation.
Common to all curves: $\tau=0.999$, $\omega_{0}=0.1\Delta$, $\eta=0.001\Delta$
and $T=0\mbox{ K}$.
Topmost curve: current in the case $\alpha=0.0$. 
}
\end{figure}

Following the procedure described in the previous section  we present in Fig.~\ref{JJ_Current:fig} the numerical results obtained for the dc Josephson
current in the presence of the EM interaction for a highly transmitting junction ($\tau=0.98$).
The dark dotted line represents the dc current $I_{DC}^{(0)}(\phi)$ obtained in absence of vibrations ($\alpha=0$) as it is
given by Eq. (\ref{eqn:8}). 
For $\omega_0=0.4 \Delta $ and $\alpha=0.04$ the dashed and solid lines show the results of the 
two-level model [Eq.\refe{eqn:11}] and the full numerical solution, respectively.
%
%
%
%
%
As anticipated before, the EM coupling induces an antiresonance on the dc Josephson current when the condition $\omega_{0}=2\omega_{A}$
is fulfilled, namely when the phase difference between L and R superconductors reaches the critical value 
\begin{eqnarray}
\phi_{res}=2\arcsin{\sqrt{\frac{1}{\tau}\big{\lbrack} 1 - (\frac{\omega_{0}}{2\Delta})^{2}\big{\rbrack} }}
\,.
\label{eqn:20} 
\end{eqnarray} 
The signature of such a modulation is the presence of dips 
in the current-phase relation, the position of which provides a measure of the vibrational frequency $\omega_{0}$ through 
the resonance condition Eq.\refe{eqn:20}. 
The width of the dip is proportional to the coupling $\alpha$ between 
the nano-resonator and the vibrational mode of the junction being excited.
It is well approximated by the expression (\ref{rabi}). 
The numerics show some additional dips in the CPR that are associated with higher-order transitions.
Such processes which are obviously absent from Eq.\refe{eqn:11}
could be incorporated by extending the analytical calculation
to the next leading orders in powers of $\alpha$ (see Refs.\cite{bergeret:2010,bergeret:2011}).
According to the upper panel of  Fig.~\ref{JJ_Current_Realistic_Parameters:fig}, the resonance dip is  accurately approximated both in position and width by the Eq.\refe{eqn:11}.
The  discrepancy  between the simplified two-level model
and the full numerical calculation becomes visible when increasing the value of $\alpha$.  This deviation is due to the fact that for large enough values of $\alpha$ the condition $\alpha<\sqrt{1-\tau}$ is not satisfied. In the particular case of the upper panel of  Fig.~\ref{JJ_Current_Realistic_Parameters:fig},  $\alpha=0.01-0.025$, $\tau=0.999$ and $\omega_{0}=0.1\Delta$.  Therefore the analytical result is valid as far as $\alpha<0.03$. 
For larger values of  $\alpha$, Eq.\refe{eqn:11} is no longer valid and one has to resort to numerical results.
These are shown in the lower panel of  Fig.~\ref{JJ_Current_Realistic_Parameters:fig}, where $\alpha = 0.04-0.1$. 
The parameters used in  Fig. 4 correspond to a high-frequency oscillator $\omega_{0}/2\pi = 1\mbox{ GHz}$
and a superconductor with a small superconducting gap $\Delta \approx 0.04 \mbox{ meV}$. Suspended carbon nanotubes between two superconductors seem to
be good candidates to reach this regime. For instance, the resonant frequency for the fundamental mode of such a vibrating nanotube was reported 
in Ref. \cite{CNT_High_Q_NEMS_SET} to be in the range of 500 MHz.  In  Ref. \cite{CNT_Cooper_Pair_Beam_Splitter}   a carbon nanotube  was connected to  a superconducting Al/Pd bilayer electrode with a BCS gap of 0.08 meV  which is in the  range of our estimation.
%
%

%
As mentioned above some differences between the numerical results and the analytical one 
emerge for $\phi$ close to $\pi$ that become more pronounced when increasing the value of the 
coupling strength $\alpha$, as shown in the lower panel of
Fig.~\ref{JJ_Current_Realistic_Parameters:fig}, for which $\alpha=0.04-0.1$. 
Notice that in this regime the exact numerical calculation predicts a change 
of the current sign close to $\phi=\pi$ due to higher order processes. 
%

The results presented show that measuring the current 
phase relation in a Josephson junction coupled to a mechanical oscillator can 
allow the detection of its periodic oscillation (for example when the oscillator 
is driven by an external force).
We found that the CPR displays sharp dips when the resonance condition $\omega_0=2\omega_A$ is met.
According to Eq.\refe{abs} for $\tau\rightarrow 1$ and $\omega_A(\phi=\pi)\rightarrow 0$,
one can in principle always satisfy the resonant condition. 
In reality this can be difficult, since it requires $\tau$ being very near to one and $\Delta$ for 
standard superconductors is much larger than the typical mechanical frequency.
This problem can be solved by using superconducting alloys with 
smaller gap, or by introducing a magnetic field in order to reduce 
$\Delta$ to a value which is slightly larger than the mechanical resonance.
Then the fine tuning of the resonance with the phase bias can be possible. 

The second crucial parameter in order to observe this effect is the 
coupling constant $\alpha$. 
A reasonable estimate of the order of magnitude of $\alpha$ can be 
obtained by considering the experiment 
of Ref.~\cite{flowers:2007}, which was performed on a driven
nanomechanical oscillator in its normal metallic state. 
The authors of Ref.~\cite{flowers:2007} estimate the resistance 
dependence on the displacement to be $(1/R) dR/dx \sim 0.1 {\rm nm}^{-1}$,
with a typical displacement in the driven case of the order of a nm.
Thus a very crude estimate of the order of magnitude of $\alpha$
is $0.1$, that is a quite strong coupling. 
Remarkably, for a suspended carbon nanotube in the Fabry-Perot regime
we find a similar order of magnitude. This can be estimated by using 
the responsivity of the transparency 
to an external change of gate voltage from Ref. \cite{kong:2001} and 
 $(1/C_g) dC_g/dx \approx 1/d$, where $C_g$ is the 
gate capacitance, and $d$ is the distance of the nanotube from the 
gate. 
It is clear that at this level these are only crude estimates of the order 
of magnitude of $\alpha$, but the result is encouraging.

In principle one could also detect thermal motion of the mechanical
oscillator with this method if the quality factor ($Q$) of the 
mechanical oscillator is sufficiently large
As a matter of facts, the thermal motion can be seen as 
a sequence of periodic oscillations with a coherence 
time given by $Q/\omega_0$. Averaging the current 
over a time much longer than this time corresponds
to averaging the current obtained above over different values 
of the amplitudes of oscillation (and thus of $\alpha$). 
One thus expects also in this case a dip in the CPR, but with 
a width that is controlled by the average of $\alpha$.
In the case of Ref. \cite{flowers:2007} this motion is tiny and gives 
that on average $\alpha \approx 10^{-4}$.
This gives an extremely sharp dip, and thus its observation 
is subject to a very accurate detection of the CPR.

\begin{figure}[ht]
\centering
	\includegraphics[width=0.5\textwidth]{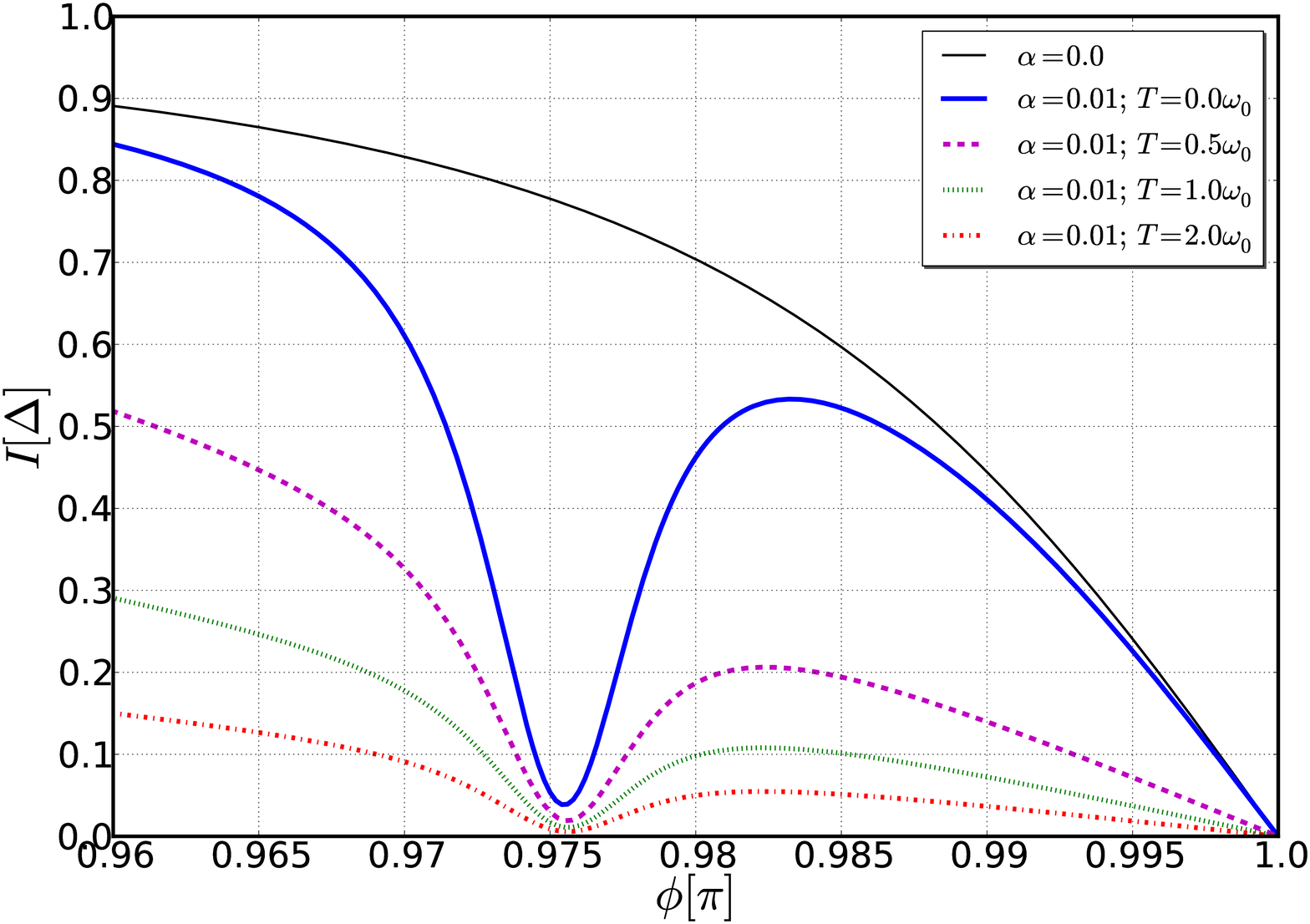}
\caption{\label{JJ_Current_Temperature} 
Josephson current in the finite temperature case $T = 0.0,0.5,1.0,2.0 \omega_{0}$, as obtained from 
the full numerical calculation. 
Parameters of the plot: $\tau=0.999$, $\omega_{0}=0.1\Delta$, $\alpha=0.01$ and $\eta=0.001\Delta$.
Topmost curve: current in the case $\alpha=0.0$. 
}
\end{figure}
%

All the results presented so far are for the zero temperature limit.
In the case of finite temperature, the coupling of the quantum 
point contact to the mechanical oscillator may lead to enhancement of the 
supercurrent as was discussed in the context
of a microwave field \cite{bergeret:2010,bergeret:2011}.
The enhancement of the current is due to inelastic
processes which promote particles from the continuum spectrum ($\omega<-\Delta$) to the lower ABS.
This phenomenon (not shown here) is an analog to the superconducting stimulation by acoustic waves in bulk materials
discussed by Eliashberg and Ivlev in 1986 \cite{eliashberg1986}.

But more important to the efficiency of this device as a detector is 
the fact that thermal fluctuations can prevent the observation
of the mechanical oscillations.
For temperatures of the order of and larger than $\omega_0=2\omega_A$, 
the thermal occupation of the two-level system tends to the value
$1/2$: $n_{\pm} = f(\pm \omega_0/2) \approx 1/2 \mp \omega_0/(8T) +o(\omega_0^3)$.
This effect will reduce the overall value of the Josephson current;
nevertheless by the exact numerical solution, we find that 
a signal is still visible till moderately high temperatures.
As it is shown in Fig.~\ref{JJ_Current_Temperature}, 
for $T=0.5-2\omega_0$ the CPRs 
maintain a local minimum at the resonance.
The position of the dip is independent of the temperature 
[cf. Eq. (\ref{eqn:20})] while it widens for higher 
temperatures.

One should emphasize that the back action of the 
Josephson junction on the  mechanical oscillator is neglected in the present work.
The establishment of an effective two-level model for the 
Josephson junction description opens the way to considering the full 
dynamics of the two systems coupled. 
At this stage, we can evaluate the ratio of the average  of the
back-action force $\langle \hat{F}_{ba}(t) \rangle = - \langle{\partial \hat{h}(t) }/{\partial x}\rangle$ to the 
elastic force $F_{el}=-m \omega_{0}^{2}x$.
We find that ${\langle \hat{F}_{ba}(t) \rangle}/{F_{el}} \approx {3\Delta}/{4m\omega_{0}^{2} \lambda^2}$,
where we introduced the characteristic length $\lambda=v/(dv/dx)$. Interestingly, this ratio is independent
of the amplitude of the oscillations.
For a single wall carbon nanotube of mass $m\approx 10^{-21} \mbox{ kg}$ that oscillates at the frequency $\omega_{0}/{2\pi}=1\mbox{ GHz}$,
we roughly estimate this ratio to be in the range $10^{-4}-10^{-6}$ \footnote{This order of magnitude for the feedback is dependent on the 
values of the $\lambda$ parameter available in the literature. We evaluated $\lambda \approx 1-10 \mbox{ nm}$ from 
Ref.\cite{flowers:2007,kong:2001}.}.
Although very small, this back action may lead to interesting effects as the cooling effect of the mechanical degree of freedom in the presence of an external magnetic field \cite{Sonne:2010}.
Note also that we have safely neglected the effects of Coulomb 
blockade on the mechanical oscillation \cite{pauget:2008,pistolesi:2007} since the 
interesting region for this device is the very transparent case.

\section{Conclusions}
\label{sec5}
In conclusion we have shown the possibility of detecting ultrafast oscillations of a nano-scale Josephson 
junction by analyzing its dc current-phase characteristics. In the high transmission regime $\tau \approx 1$, 
the onset of electromechanical coupling results in the appearance of dips in the $I_{DC}(\phi)$ characteristics
for precise values of the phase difference $\phi$. The location of those dips provides a new way to measure the
vibrational frequency $\omega_{0}$ of the oscillator, and their width 
is directly proportional to the EM coupling strength $\alpha$. If the latter is sufficiently small,
we have derived an effective two-level Hamiltonian [Eq.\refe{tlha}] which describes quite accurately 
the dynamics of the contact in the presence of an EM perturbation.  
Our results provide a new way of characterizing the motion at the nanoscale.

\section*{Acknowledgements}
\label{secack}
The authors are grateful to Bruno Rousseau for precious help with Matplotlib and to Francois Lefloch for useful
correspondance. F. S. B. acknowledges the Spanish MICINN
(Contract No.\ FIS2008-04209), the
CSIC (Intramural Project No.\ 200960I036), and the Basque Government under UPV/EHU Project IT-366-07 for financial support.
F.P. acknowledges the French Agence Nationale Recherche for finanancial support (Contract QNM ANR10-BLAN-0404-03).

\section*{Appendix : Free Green functions}
\label{secAppendix}

Here we determine the GFs of the nonvibrating Josephson junction ($\alpha=0$).
We first consider the isolated Right electrode described by the 
time-independent Hamiltonian (Eq.\refe{eqn:3}).   
The retarded (R) and advanced (A) surface Green functions of such a system can be found analytically 
by making use of the periodicity of the Hamiltonian when writing its Dyson equation  

\begin{eqnarray}
\hat{g}_{R}^{\eta=R(A)}(\omega) = \Big{\lbrace} 
\omega_{\eta} -\Delta \hat{\sigma}_{x}
- \hat{v}_{0}\hat{g}_{R}^{\eta=R(A)}(\omega)\hat{v}^{\dagger}_{0}\Big{\rbrace}^{-1}
\label{eqn:21} 
\end{eqnarray}

\noindent
where 
$\omega_{\eta}=\omega + i\eta$. The small imaginary part describes inelastic scattering in the leads
within the relaxation time approximation 
and is the smallest energy scale of the problem.
In the small gap limit ($\Delta, \omega << 2|v_{0}|$, $v_0$ is the electrode bandwidth), one
finds for the solution of Eq.\ref{eqn:21}

\begin{eqnarray}
\hat{g}_{S}^{\eta=R(A)}(\omega) = N_{\eta}(\omega)\Big{\lbrace} 
\omega_{\eta} + \Delta \hat{\sigma}_{x}\Big{\rbrace} 
\label{eqn:22} 
\end{eqnarray}

\noindent
where $N_{\eta}(\omega)=-\frac{1}{v_{0}} \Big{\lbrace} \frac{\theta(|\Delta|-|\omega|)}{\sqrt{|\Delta|^2-\omega_{\eta}^2}} + i\eta \mbox{sign}(\omega) \frac{\theta(|\omega|-|\Delta|)}{\sqrt{\omega_{\eta}^2-|\Delta|^2}} \Big{\rbrace}$. \\

\indent
We connect now the R-lead to the L-lead through the time-independent part of the tunnel Hamiltonian (Eq.\ref{eqn:4}). 
The Dyson equations for the retarded (advanced) GFs of the entire nano-junction read

\begin{eqnarray}
\hat{G}_{R}^{\eta=R(A)}(\omega) &=& \frac{1}{\big{(}\hat{g}_{R}^{\eta=R(A)} \big{)}^{-1}(\omega) - \hat{v}^{\dagger}\hat{g}_{L}^{\eta=R(A)}(\omega)\hat{v} }  \label{eqn:23}  \\
\hat{G}_{LR}^{\eta=R(A)}(\omega) &=& \hat{g}_{L}^{\eta=R(A)}(\omega)\hat{v}\hat{G}_{R}^{\eta=R(A)}(\omega) \label{eqn:24} 
\end{eqnarray}

\noindent
Eqs.(\ref{eqn:23}-\ref{eqn:24}) can be solved analytically and provide the expressions for the 
unperturbed GFs used in Eq.\refe{dyson} 

\begin{eqnarray} 
\hat{G}_{R}^{\eta=R,A}(\omega) &=& \frac{g_{\eta}(\omega)}{|v_{0}|(1+\beta)(\omega_{\eta}^2-\omega_{A}^2)}
\left[ \begin{array}{cc}
\omega_{\eta} & \underline{\omega}_{A}^{*}  \label{eqn:25}  \ \\
\underline{\omega}_{A} & \omega_{\eta} \end{array} \right]
\\
\hat{G}_{LR}^{\eta=R,A}(\omega) &=& \frac{\beta}{|v|(1+\beta)(\omega_{\eta}^2-\omega_{A}^2)}
\left[ \begin{array}{cc}
a(\phi) & -b(-\phi) \label{eqn:26}  \ \\
b(\phi) & -a(-\phi) \end{array} \right]
\nonumber\\
a(\phi) &=& e^{i\phi/2}\omega_{\eta}^{2}-e^{-i\phi/2}\underline{\omega}_{A}\Delta 
\\
b(\phi) &=& \omega_{\eta}(e^{i\phi/2}\Delta-e^{-\phi/2}\underline{\omega}_{A})
\end{eqnarray} 

\noindent
where we introduced the intermediate function $g_{\eta}(\omega)=\theta(|\Delta|-|\omega|)\sqrt{|\Delta|^2
-\omega_{\eta}^2}- i\eta \mbox{sign}(\omega)\theta(|\omega|-|\Delta|)\sqrt{\omega_{\eta}^2-|\Delta|^2}$ and the 
complex number
$\underline{\omega}_{A}=\Delta\frac{1+\beta e^{i\phi}}{1+\beta}$, the module of which is the Andreev bound 
state energy $\omega_{A}=\Delta\sqrt{1-\tau\sin^{2}(\phi/2)}$. The remaining components of the GFs are obtained by changing 
simultaneously left and right electrode indexes and the sign of the superconducting phase difference,
i.e., $(L,\phi)\rightarrow (R,-\phi)$. \\

\noindent
Finally, the non diagonal component of the Keldysh GFs can be found using the relation (valid at equilibrium only) 

\begin{eqnarray} 
\hat{G}_{XX'}^{+-}(\omega)=f(\omega) \big{\lbrace} \hat{G}_{XX'}^{A}(\omega) - \hat{G}_{XX'}^{R}(\omega) \big{\rbrace}\;,
\label{eqn:27} 
\end{eqnarray} 

\noindent
where $f$ is the Fermi distribution function.
From this equation one can compute the equilibrium Josephson current (Eq.\ref{eqn:8}) by integrating the
following expression 

\begin{eqnarray}
I_{DC}^{(0)}(\phi) = \int \frac{d\omega}{2\pi} \mbox{tr} \Big{\lbrace} \hat{\sigma}_{z} \Big{\lbrack}
\hat{v}\hat{G}_{RL}^{+-}(\omega) - \hat{v}^{\dagger}\hat{G}_{LR}^{+-}(\omega) \Big{\rbrack} \Big{\rbrace}
\,.
\label{eqn:28} 
\end{eqnarray}

\bibliography{biblioNEMS}

\end{document}